\documentclass[10pt,conference]{IEEEtran}

\usepackage[utf8]{inputenc}
\usepackage{xcolor, soul}
\usepackage[normalem]{ulem}
\usepackage{bm}
\usepackage{booktabs}
\usepackage{hyperref}
\usepackage{enumitem}
\usepackage{subcaption,graphicx}

\IEEEoverridecommandlockouts
\usepackage{cite}
\usepackage{amsmath,amssymb,amsfonts}
\usepackage{algorithmic}
\usepackage{graphicx}

\usepackage{tikz}
\definecolor{pink}{rgb}{0.9,0,0.9}

\begin{document}

\title{Issue Link Label Recovery and Prediction for Open Source Software}

\author{
\IEEEauthorblockN{Alexander Nicholson }
\IEEEauthorblockA{\textit{School of Computer Science} \\
\textit{McGill University}\\
Montreal, Canada \\
alexander.nicholson@mail.mcgill.ca}
\and
\IEEEauthorblockN{ Jin L.C. Guo}
\IEEEauthorblockA{\textit{School of Computer Science} \\
\textit{McGill University}\\
Montreal, Canada \\
jguo@cs.mcgill.ca}
}

\maketitle

\begin{abstract}
Modern open source software development heavily relies on the issue tracking systems to manage their feature requests, bug reports, tasks, and other similar artifacts. Together, those ``issues'' form a complex network with links to each other. The heterogeneous character of issues inherently results in varied link types and therefore poses a great challenge for users to create and maintain the label of the link manually. The goal of most existing automated issue link construction techniques ceases with only examining the existence of links between issues. In this work, we focus on the next important question of whether we can assess the type of issue link automatically through a data-driven method. We analyze the links between issues and their labels used the issue tracking system for 66 open source projects. Using three projects, we demonstrate promising results when using supervised machine learning classification for the task of link label recovery with careful model selection and tuning, achieving F1 scores of between 0.56-0.70 for the three studied projects. Further, the performance of our method for future link label prediction is convincing when there is sufficient historical data. Our work signifies the first step in systematically manage and maintain issue links faced in practice.

\end{abstract}


\section{Introduction}
\label{sec:intro}



\begin{figure}[t!]
    \centering
    \includegraphics[width=0.45\textwidth]{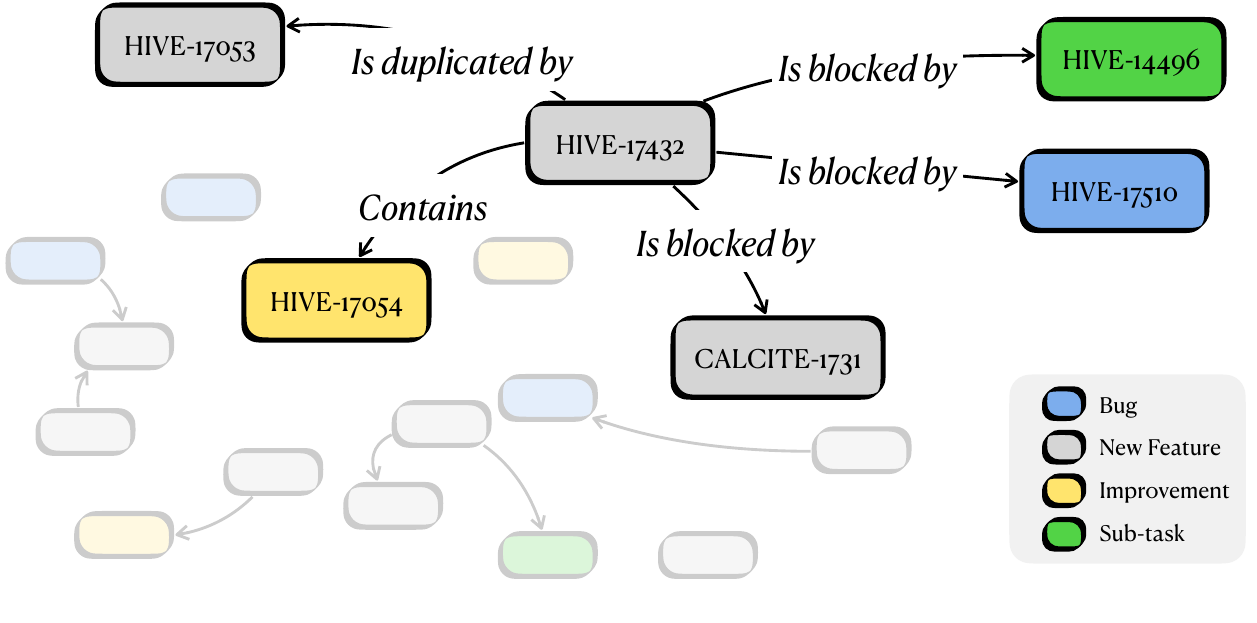}  
 \caption{Example of an issue (\texttt{HIVE-17432}) and different type of horizontal trace links linked to this issue in Jira. }
 \label{fig:issue_example}
 \vspace{-6pt}
\end{figure}

Issue tracking systems (ITSs) play an increasingly important role in modern open source software development as a central location for submitting and managing software \textit{issue}s. An \textit{issue} in this context often represents a piece of information about the current system or a unit of work to be completed to improve the software product. This could be a requested feature, a reported bug, a planned task, missing documentation, etc~\cite{da2016impact}. In practice, depending on the complexity of software projects and how the projects are managed, meaningful associations often form among issues, referred to as \textbf{horizontal trace links}~\cite{heck2014horizontal}. Figure \ref{fig:issue_example} exemplifies the local connections of one issue from one of the open source projects in our case study HIVE\footnote{hive.apache.org} using Atlassian Jira\footnote{https://www.atlassian.com/software/jira}. This feature request is linked to two other features, one sub-task, and one bug.

The heterogeneous nature of the content of the issues in the ITSs leads to substantial variance in the \textbf{types} of the trace links between issues. To record the specific type for links, Jira allows the developer to assign particular labels as metadata of the links.  In Figure \ref{fig:issue_example}, the issue \texttt{HIVE-17432} was linked with other issues using three distinct labels, i.e. \textit{contains}, \textit{is blocked by}, \textit{is duplicated by}. Software stakeholders rely on this information to identify all the tightly related issues and their particular relationships to perform different tasks such as prioritizing the new features, understanding the current status of development, communicating the progress towards completing certain goals~\cite{da2016impact, arya2019analysis}, just to name a few.  

Making use of the link label functionality in the ITSs requires additional effort to create and maintain these labels~\cite{nicholson2020traceability}. Currently, project stakeholders can only assign and update the labels manually. Given that the experience and preference of the stakeholders vary considerably in open source projects, the same link labels might be used inconsistently within and across projects. Moreover, when the issues evolve over time, extra work is devoted to reevaluate the link labels or they will become outdated or unreliable. Therefore, automated support for suggesting and updating the link labels is a non-negligible step for enhancing the current ITSs to achieve effective issue navigation~\cite{Tomova:2018:UTL:3183440.3195086} and reasoning~\cite{luders2019openreq}.

In this work, we take on the task of \textit{automated recovery and prediction of labels for horizontal trace links in the ITSs}. Link label recovery refers to the scenario of inferring link labels for issues that have already been in the ITS, while Issue link label prediction refers to predicting link labels for future (i.e. unseen) issues. By using the existing trace link labels and machine learning techniques, we establish a method for recovering and predicting trace link labels that can adapt to the nuances of different software projects. Furthermore, we incorporate knowledge external to improve the learning of trace link labels. The performance of our methods is evaluated in two different scenarios to simulate the context of link label recovery and prediction in practice, respectively. In particular, we address the following key research questions: 

\textbf{RQ1}: What are the characteristics of labels used for issue links in open source projects? 

\textbf{RQ2}: To what extent can we recover link labels using machine learning techniques combining the textual content of issues, textual content external to the ITS and the context information about the issues and issue links?

\textbf{RQ3}: To what extent can we recommend the labels for future links giving the historical link labels in the project? 

We observe that the link labels are heavily used in the ITSs for trace links between issues. Our method can effectively recover link labels in the ITS for various sized projects. Given a sufficient amount of training data, our method can further predict labels for future links. The contribution of this work is two-fold. First, our work complements the existing research on horizontal traceability that mainly focused on constructing trace links without considering the particular types of links. Second, we propose a novel link label recovery and prediction solution that can reasonably fit into the state-of-the-art ITSs. 


\section{Related Work}
\label{sec:related_work}

\subsection{Software Traceability}
Previous work on software traceability has ranged from recovering and analyzing, to interpreting and making use of artifact connections. Several recent efforts has focused on analyzing the connections between various artifacts written informally in terms of language use, including code reviews \cite{hirao2019review}, question and answer \cite{fu2017easy} and bug reports \cite{hu2018recommending}, among others. Our work is closer to this line of work. Rather than choosing one particular link or artifact type, we focus on the breadth of link types frequently used in ITS.

Researchers and practitioners have additionally investigated the use of trace links to support a wide variety of development and maintenance tasks such as release planning \cite{Dahlstedt2005}, impact analysis \cite{davide_TSE2018}, completion analysis, source-code justification analysis \cite{Rempel2017PreventingDT}, test result and fault tracing \cite{Sthl2016AchievingTI} and compliance verification \cite{Mder2013StrategicTF}. The link label recovery and prediction solution proposed in this work can be used as a precursor to any of the above tasks that rely on certain types of links.

\subsection{Automated Trace Link Construction}

Given the magnitude of the traceability problem, there has been much previous work on automated and semi-automated solutions to trace link creation and maintenance \cite{Guo:2014:TID:2642937.2642970, guo2017semantically, abukwaik2018semi}. The majority of this work deploys information retrieval or machine learning techniques but does not handle suggesting link types during trace link creation. There has also been work on heuristic-based trace link creation methods \cite{guo_re2013, spanoudakis2004rule}. In particular, the links can be grouped into different categories based on which heuristic has been applied to create the link \cite{spanoudakis2004rule}. Instead of a heuristic-focused approach, we investigate automated solutions to the problem of trace link label recovery and prediction for the ITS in which the historical labels can be mined and used for training the machine learning models.

\subsection{Trace links in Issue Tracking Systems.}


ITSs resemble change requests or just-in-time requirements that are triaged and addressed by team members continually~\cite{Do2017RefinementAR}. As a result, traceability for ITSs has attracted some attention in recent literature \cite{nicholson2020traceability, Tomova:2018:UTL:3183440.3195086}. The relationship between on particular issue has been investigated in previous work. For example, Maalej, et. al studied the links between \textit{tasks} \cite{maalej2017using} for the purpose of work planning. Tasks, however, only represent a marginal segment of a project's software artifacts (see Section~\ref{sec:dataset}). Similarly, different link types have been considered between question and answer \cite{fu2017easy} and bug reports \cite{hu2018recommending}. Our initial observation reveals that the horizontal links between issues are often across issue types (e.g. see Figure~\ref{fig:issue_example}). Therefore, we expand the scope of link types across software issues that represent a rich set of artifact types such as bug reports, feature requests, improvements, etc.  


\subsection{Issue Trace link Types} 
In the context of investigating link labels used for open source projects, our work is closely related to the recent studies by Tomova et. al.~\cite{Tomova:2018:UTL:3183440.3195086} and Nicholson et. al~\cite{icse2018_traceability_in_the_wild}. Tomova et. al. presented preliminary results on recovering link labels on seven OSS projects using Jira~\cite{Tomova:2018:UTL:3183440.3195086}. They focused on four default link labels provided by Jira and discussed the impact of selecting threshold when using a term matching based method. Their observation demonstrates the ineffective for retrieving links with the type of ``relates to'' and calls for closer investigation on link label usage in OSS and tool support. Nicholson et. al. provided a more comprehensive picture of how the link labels were used across 66 OSS projects using Jira~\cite{Tomova:2018:UTL:3183440.3195086}. Their study reveals that the customized link labels are commonly used, and can appear more frequently than the default label of ``is clone of''. They also demonstrated that the network structure formed by historical links and their labels can be useful to extract patterns to recover missing links and corresponding labels. The dependency of one particular network structure, however, limits the applicability of such a method on predicting the label of a wide set of links.

\section{Issue Link Characteristics and Dataset Creation}
\label{sec:dataset}
In this section, we perform a series of empirical investigations to understand the explicit links that project stakeholders assign between issues (RQ1). We further describe the steps that we have taken to prepare the selected project datasets for the experiments of automated link label recovery (in Section~\ref{sec:semantics_recovery}) and prediction (in Section~\ref{sec:semantics_prediction}). 

\subsection{Issue Links and Their Labels for Apache Projects}
\label{subsec:all_projects}

Our study is based on a dataset created by Nicholson et. al. for their empirical study of Jira issue networks~\cite{alexander_nicholson_link_dataset, nicholson2020traceability}. The complete dataset was mined from the sixty-six Apache Open Source projects and contains a total of $71,087$ trace links between $90,936$ issues. To the best of our knowledge, this is the largest dataset on horizontal trace links. It also contains a rich set of metadata including the link labels that represent the particular types specified by issue creators or maintainers.

As shown in Figure~\ref{fig:issue_example}, Jira allows users to assign distinct labels to issue links to indicate different types of links. Jira provides four default labels: \textit{relates to / relates to}, \textit{duplicates / is duplicated by}, \textit{blocks / is blocked by} and \textit{clones / is cloned by}\footnote{https://confluence.atlassian.com/adminjiraserver073/configuring-issue-linking-861253998.html}. Authorized project stakeholders can also create and customize new labels to express additional associations between issues according to their needs. The labels saved in the ITS serve as a lens through which we can examine the issue link types on a large scale.  Note that although Jira permits linking issues across different projects, in this work, we only focus on within-project links that constitute a sizable portion of the links in the ITSs.

\begin{figure}
    \centering
    \includegraphics[width=0.40\textwidth]{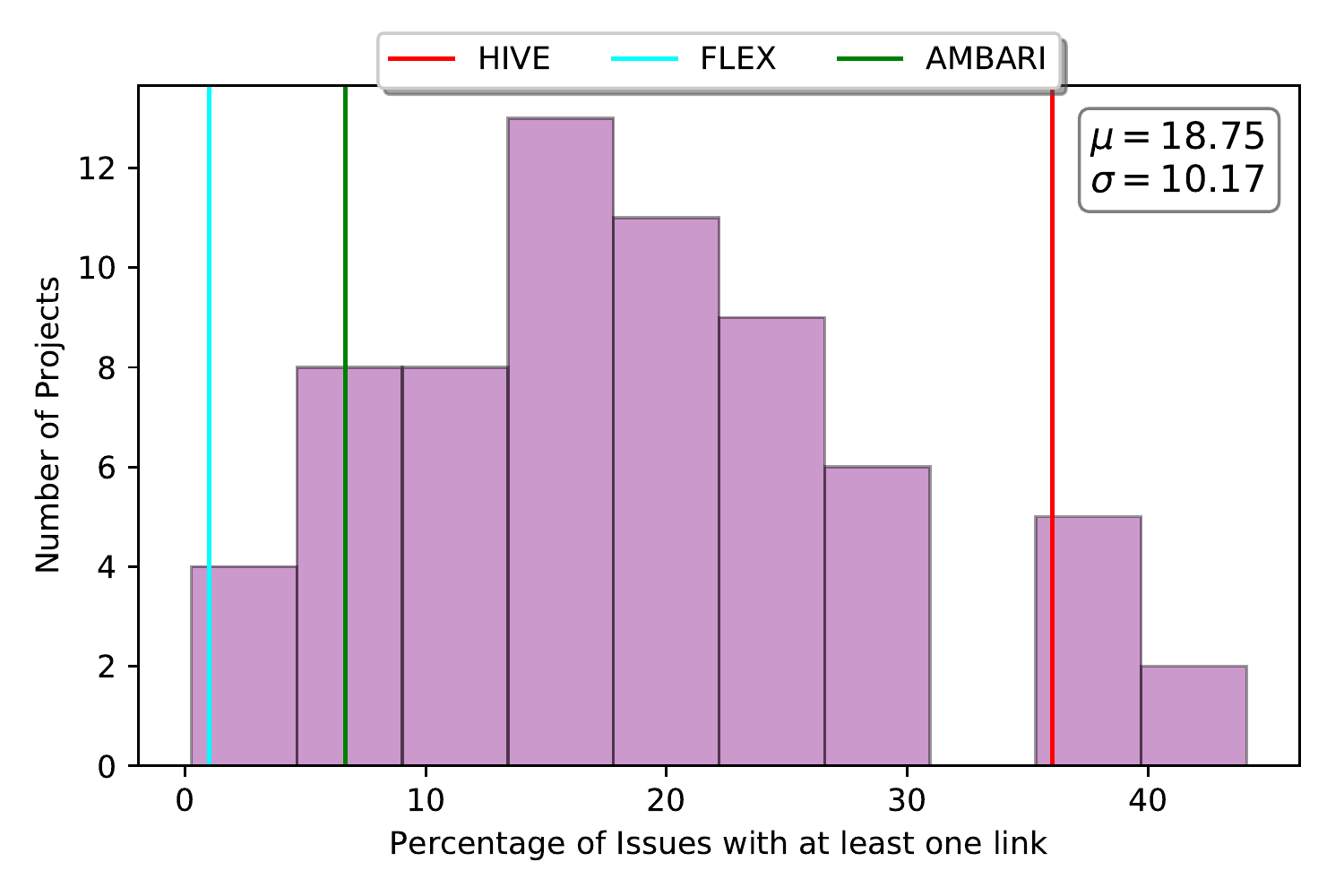}
    \caption{Distribution histogram of the ratio of linked issues to total issues in 66 Apache projects~\cite{alexander_nicholson_link_dataset}. The location of the case study projects is indicated by the vertical lines. }
    \label{fig:linked_issues_percentage}
\end{figure}

We first examine how prevalent the issues contain horizontal trace links in each project. To achieve this, we calculate the ratio of issues that have at least one link to the total number of issues in that project. Figure \ref{fig:linked_issues_percentage} shows the histogram of this ratio across all projects. On average, more than 18\% of issues in a project have links to other issues with a standard deviation of 10\%. In terms of the link labels used in practice, previous work dataset suggests that the distribution of the occurrence of different labels is highly skewed~\cite{nicholson2020traceability} -- among 16 unique link labels, \textit{related to} label accounts for nearly half of the links (45.56\%). Furthermore, a non-trivial portion of link label categories falls below the $1\%$ of the trace links. 

\subsection{Project Case Studies and Dataset Creation}
\label{subsec:case_study_dataset}

To further investigate the link labels used in practice and the potential of using machine learning techniques to alleviate the manual effort, we focus on three projects from this dataset~\cite{alexander_nicholson_link_dataset} for the remaining study, i.e. Hive, Ambari, and Flex \footnote{http://\{hive/ambari/flex\}.apache.org}. This selection is based on the following important reasons: First, all three projects are actively being developed and maintained. Second, they represent projects with different characteristics in terms of issue traceability (see Figure~\ref{fig:linked_issues_percentage}). Third, they have active CWiki pages which are the project-specific wikis managed by the project community. CWiki pages contain supplementary project information such as release notes, documentation and discussions which enable us to incorporate additional project specific data during link label recovery (further discussed in Section~\ref{subsec:issue_encoding}). Among all the 66 projects we have examined, these three projects are the only ones that we can obtain such project documentation with reasonable effort.

The breakdown of issue types that contain links in each studied project is summarized in Table~\ref{table:typebreakdown}. We also present the link labels with their respective quantities in Table~\ref{table:linkbreakdown}. Except for \textit{bugs} and \textit{tasks}, the issue type of \textit{improvement} also frequently contains links to other issues, ranging between 10.45\% to 17.61\%. Furthermore, we observe that 14.17\% - 33.85\% of the total links are between \textit{bug} and \textit{non-bug} issues. This big portion of links has been omitted from the previous work that only focuses on links between one issue types~\cite{yang2016combining, xia2015elblocker}. Additionally, the overlapping between the less frequent labels with the more dominant ones is also observed in these datasets, such as \textit{dependent} and \textit{depends upon}, \textit{Blocked} and \textit{blocks}.  These overlapping might pose risk to the quality of the dataset. To reduce such a potential threat to the validity of our study, we exclude the link labels if their occurrences account for less than $1\%$ of the total links in that project and if the occurrences are less than 20 in our link semantic solution.



\begin{table}[t!]
\centering
\caption{The types of issues that contain links and their occurrence in project AMBARI, FLEX, and HIVE respectively. The types are ranked based on their occurrence in project AMBARI and the occurrence with more than 10\% are highlighted.}
\setlength\tabcolsep{4pt}
\begin{tabular}{c|ccc}
\toprule
    Issue Type           & AMBARI                  & FLEX                    & HIVE                    \\ 
\midrule

Bug   & \textbf{983 / 65.01\%} & \textbf{267 / 73.76\%} & \textbf{3682 / 54.71\%} \\
Task           & \textbf{228 / 15.08\%}          & 21 / 5.80\%           & 205 / 3.05\%         \\
Improvement    & \textbf{158 / 10.45\%}           & \textbf{45 / 12.43\%}           & \textbf{1185 / 17.61\%}         \\ 
New Feature    & 54 / 3.57\%            & 12 / 3.32\%            & 454 / 6.75\%           \\ 
Epic           & 41 / 2.71\%            & 0 / 0.0\%               & 0 / 0.0\%               \\
Sub-task       & 20 / 1.32\%            & 12 / 3.32\%            & \textbf{1105 / 16.42\%}        \\ 
Technical task & 12 / 0.79\%            & 0 / 0.0\%               & 0 / 0.0\%               \\ 
Story          & 11 / 0.73\%            & 0 / 0.0\%               & 0 / 0.0\%               \\
Documentation  & 3 / 0.20\%             & 0 / 0.0\%               & 0 / 0.0\%               \\ 
Test           & 1 / 0.07\%             & 0 / 0.0\%               & 85 / 1.26\%            \\ 
Wish           & 1 / 0.07\%             & 4 / 1.11\%             & 14 / 0.21\%            \\ 
Question       & 0 / 0.0\%               & 1 / 0.28\%             & 0 / 0.0\%               \\ \hline 
Total           & 1512             & 362             & 6730            \\
\bottomrule
\end{tabular}\\
\label{table:typebreakdown}
\end{table}



\begin{table}[t!]
\centering
\caption{Link labels and their occurrence in project AMBARI, FLEX, and HIVE respectively, with more than 10\% are highlighted. Note, the labels with percentage $< 1\%$ or occurrence $< 20$ are excluded from our case study for that project.}
\begin{tabular}{c|ccc}
\toprule
  Link Label             & AMBARI                  & FLEX                    & HIVE                    \\ 
\midrule
relates to     & \textbf{310 / 32.91}\% & \textbf{94 / 38.06}\% & \textbf{3060 / 52.66}\% \\ 
duplicates     & \textbf{305 / 32.38}\% & \textbf{51 / 20.65}\% & \textbf{708 / 12.18}\%  \\ 
blocks         & 89 / 9.45\%   & 20 / 8.10\%  & \textbf{717 / 12.34}\%  \\ 
depends upon   & 70 / 7.43\%   & 13 / 5.26\%  & 373 / 6.42\%   \\ 
requires       & 38 / 4.03\%   & 20 / 8.10\%  & 134 / 2.31\%   \\ 
contains       & 27 / 2.87\%   & 2 / 0.81\%    & 103 / 1.77\%   \\
is a clone of  & 27 / 2.87\%   & 23 / 9.31\%  & 71 / 1.22\%    \\ 
breaks         & 26 / 2.76\%    & 14 / 5.67\%  & 190 / 3.27\%    \\
incorporates   & 21 / 2.23\%   & 8 / 3.24\%   & 339 / 5.83\%   \\ 
supercedes     & 15 / 1.60\%   & 2 / 0.81\%    & 84 / 1.45\%    \\ 
causes         & 6 / 0.64\%    & 0 / 0.0\%     & 11 / 0.19\%    \\ 
Blocked        & 5 / 0.53\%    & 0 / 0.0\%     & 11 / 0.19\%    \\ 
is a parent of & 2 / 0.21\%    & 0 / 0.0\%     & 3 / 0.05\%     \\
Dependent      & 1 / 0.11\%    & 0 / 0.0\%     & 5 / 0.09\%     \\ 
Dependency     & 0 / 0.0\%      & 0 / 0.0\%     & 1 / 0.02\%     \\ 
Parent Feature & 0 / 0.0\%      & 0 / 0.0\%     & 1 / 0.02\%     \\ \hline
Total          & 942            & 247           & 5811            \\
\bottomrule
\end{tabular}\\
\label{table:linkbreakdown}
\end{table}

When creating the dataset for the remaining study, i.e. automated link label recovery and prediction, we retrieve issue summaries, descriptions, types, resolution status, linked issues, as well as link labels. Among them, \textit{summary} contains a brief outline of the issue while the \textit{description} details the technicalities of the issue such as the reproduction steps in the case of a bug, or the detailed requirements in the case of a feature request. In our dataset, the \textit{summary} lengths range from an average of $57$ characters in Hive to $65$ characters in Flex. \textit{Descriptions} are, as anticipated, more detailed, containing an average of from $863$ characters in Ambari to $1059$ characters in Flex. 

When creating the dataset for model training and testing, we follow a standard text pre-processing method to normalize both the issue summary and description: we first remove all non-alphanumeric characters and converted the remaining characters to lowercase; we then remove English stop words from the remaining tokens using the NLTK Python library\footnote{http://www.nltk.org/}; for known words, we then convert them to their standard lemma form. Additionally, camel-cased tokens are split into separate tokens using regular expressions and also converted to lowercase and their standard lemma form.

\section{Issue Link Representation}
\label{sec:semantics_features}


In this section, we seek to represent the issue links for recovering and predicting their labels in practice using machine learning techniques. We first discuss how we extract the textual features from the issue content. We then extend the textual features to incorporate additional resources external to the ITS. Finally, we use the metadata of issues to capture the context in the ITS. We use concatenated feature sets as input to recover and predict link labels in the next two sections.

   
   
    
    
    
    
%

\subsection{Text Encoding using Issue Content}
\label{subsec:tfidf_encoding}
Textual features are heavily used in the literature to assess the existence of trace link~\cite{heck2014horizontal, guo2017tackling}. Term Frequency-Inverse Document Frequency (TF-IDF) model, one of the most common ways to make decisions on document relevancy, serves as a strong baseline when analyzing issue content~\cite{wu2008interpreting_tfidf, arya2019analysis}. Intuitively, TF-IDF model capture how important a word is to each document in a collection of documents. We use the Scikit-Learn library~\cite{scikit-learn} to transfer each pre-processed text into the vector representation.

Recall that the summary and description of issues contain information with different granularity. To retain this distinction while capturing the textual content of each project, summaries and descriptions are treated as separate documents when fitting the TF-IDF model. The resulting issue vector then is the concatenation of the vector representation of its summary and description which produces a vector of size $2 \times N$ where  $N$ is the vocabulary size of the project. Each link is then encoded as a fixed-sized feature vector concatenating the TF-IDF representation of two issues in question. The textural features generated this way is denoted by $textEnc_{tfidf}$.

\subsection{Text Encoding Integrating External Resources}
\label{subsec:issue_encoding}

Depending on the granularity of the issues and the terminology used by the issue authors, the gap of the textural content between two linked issues can be significantly large~\cite{guo2017tackling}. This gap can be potentially filled with information about the language usage, the application domain, and the project itself. To consolidate different kinds of information for link labels related tasks, we investigate alternative text encoding functions with word embedding techniques. Different from the TF-IDF model, word embedding refers to the process of transforming continuous vector representations of words from a high-dimensional space to one with much fewer dimensions~\cite{mikolov2013efficient}. It can capture several important syntactic and semantic properties of the words when trained to build a language model using a large collection of documents~\cite{landauer1997solution}. Using word embeddings as input has been shown to greatly improve performance in many natural language processing related tasks such as image captioning~\cite{karpathy2015deep} and sentiment analysis~\cite{nakov2016semeval}. In the context of software requirement analysis, it also has been applied tasks such as identification of ambiguous cross-domain terms~\cite{mishra2019use}.


We primarily take advantage of the large amount of online text about general concepts, software development and specific open source projects to pre-train the word embeddings to capture the distribution of word usage in different corpus. The word embeddings can then be updated to fit the issue text for each project. Particularly, the following resources are used in our study, with an increasing level of relevance to the domain:

\noindent $\bullet$ \textbf{Wikipedia (Wiki)}: We use the wikitext103 dataset \cite{DBLP:journals/corr/MerityXBS16} which consists of verified Wikipedia articles with a total of over 103 million words in size. Previous work has demonstrated that word embeddings trained on this dataset can effectively capture the statistical distribution of general terms.

\noindent $\bullet$ \textbf{Stack Overflow (SO)}: We use the StackSample dataset\footnote{https://www.kaggle.com/stackoverflow/stacksample} which contained $10\%$ of Stack Overflow questions and answers as of 2016. 
We include this dataset because the previous study has suggested that software-specific documents might be more effective than general-purpose corpus to capture word similarities for tasks that rely on processing software artifacts~\cite{Tian:2014:SSW:2591062.2591071}.
    
\noindent $\bullet$ \textbf{Project Documentation (PD)}: Cwiki pages of each project in our case study\footnote{https://cwiki.apache.org/confluence/collector/pages.action? \newline key=\{FLEX/Ambari/Hive\}} contain the information such as documentation, release notes, announcements, reference guides, and frequently asked questions. These were mined using the selenium webdriver tool\footnote{https://www.seleniumhq.org/projects/webdriver/}. We aim to explore the potential of capturing project specific context for link label recovery through word embedding techniques. We mined a total of $247$ pages from the Ambari project, $327$ from the Flex project and $57$ from the Hive project.

We represent each word in the dataset vocabulary as a fixed-sized dimensional fastText embedding~\cite{fastText}. The fastText embedding type was chosen because it additionally accounts for the co-occurrence of sub-word tokens (i.e. groups of characters), and introduces various optimizations for training speed. To aggregate a document's word vectors we average the word embedding vectors, which is a technique known as a Bag of Vectors \cite{P18-1198}.

Once we have used the document collections above to train a set of word embeddings (\textit{\{wiki, stack, proj\}}), we update the embeddings by training using the same objective on entire issue text (summaries and descriptions) from the respective project. The word embedding text encoding techniques pre-trained by Wikipedia, Stack Overflow, and project documentation are denoted by $textEnc_{wiki}$, $textEnc_{stack}$ and $textEnc_{proj}$ respectively.


\subsection{Issue Metadata}
\label{subsec:metadata}

Metadata from the ITSs has proven to be useful to suggest trace links between commit message and issue post~\cite{icse2018_traceability_in_the_wild}. We hypothesize that  
the reason for assigning certain link labels is also correlated with the contextual information that can be extracted through the issue metadata. An example of one typical issue from Jira is given in Figure~\ref{fig:issue_example}. In particular, we include the following metadata when encoding the issue links:

\noindent $\bullet$ \textbf{$timeDelta(issue_1, issue_2)$}: The normalize difference between the creation time (in days) of $issue_1$, and $issue_2$ respectively;

\noindent $\bullet$ \textbf{$type(issue)$}: The issue type (see Table \ref{table:typebreakdown});

\noindent $\bullet$ \textbf{$assignee(issue)$}: The unique identifier of the issue assignee. An special identifier is used for no assignee.

\noindent $\bullet$ \textbf{$reporter(issue)$}: The unique identifier of the issue reporter.

This metadata is then transformed into a feature-set. The categorical metadata items such as type, assignee, and reporter are all represented as $N$-dimensional one-hot vectors, where $N$ is the number of categories, that contains $0$ in all indexes except the index corresponding to the category in question which contains $1$.


\section{Automated Trace Link Label Recovery}
\label{sec:semantics_recovery}
\begin{figure}[!b]
    \centering
    \includegraphics[width=0.45\textwidth]{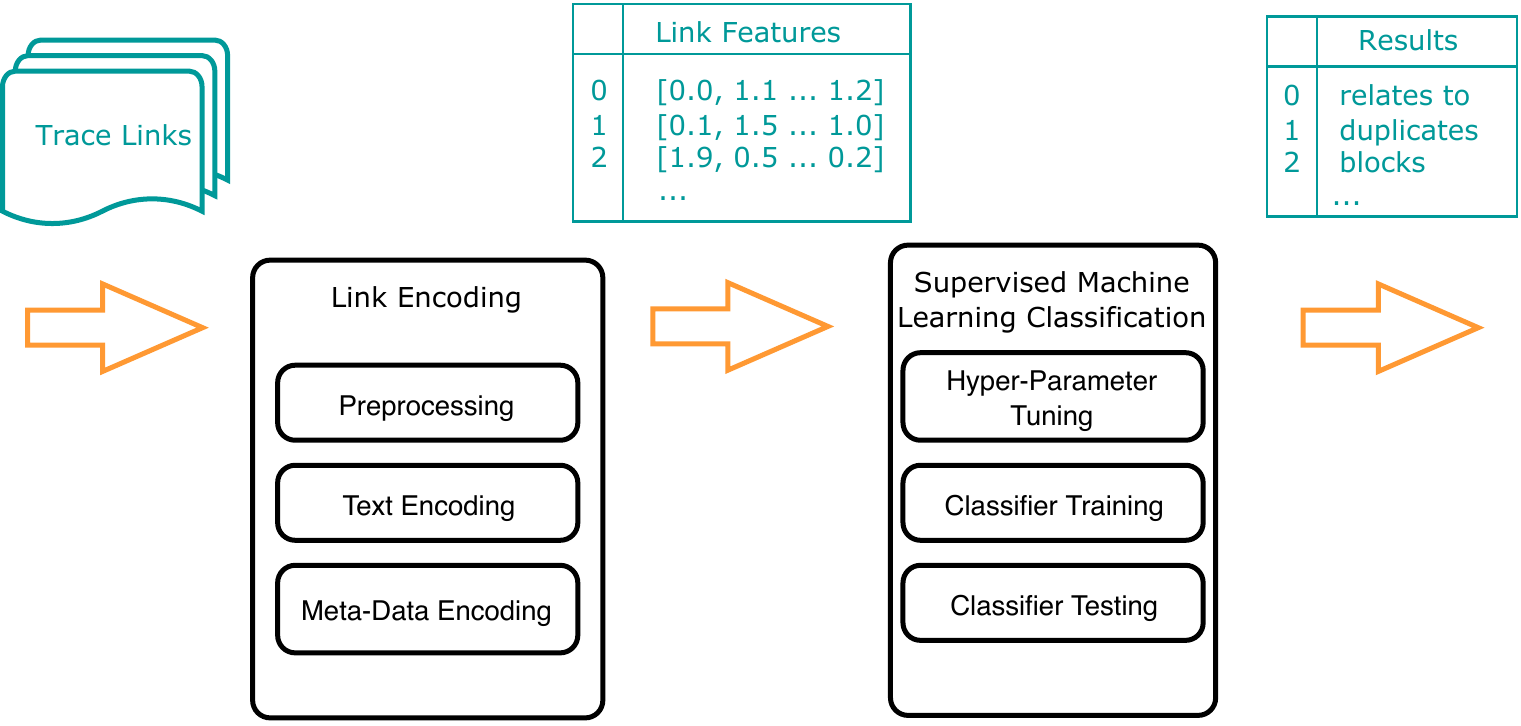}
    \caption{The complete processes of classifying the link labels.}
    \label{fig:process_pipeline}
\end{figure}

Automated issue  trace  link  solutions aims to assist  the  tasks such  as  the recovery  of  missing  link  labels  and  link  label  maintenance. Given a certain amount of labels provided by the project stakeholders,such solutions improve the quality of the remaining labels between issues saved in the ITS. This process can serve to support project management as well as other retrieval based tasks. Towards this end, we can build statistical classifiers that learn from the important features of the existing link labels and suggest the labels for other links under consideration. We depict the complete process of our method in Figure~\ref{fig:process_pipeline}.

\subsection{Supervised Machine Learning Classification}
\label{subsec:mlclassifoverview}

In this work we experiment with the following supervised machine learning models due to their wide adoption in text mining tasks \cite{zhang2006multilabel} and the improved results reported in recent work on software artifacts \cite{scandariato2014predicting, arya2019analysis}.

\noindent $\bullet$ \textbf{Logistic Regression (\textit{LR})}: A parametric model which fits parameters to minimize the cross-entropy between the model predictions and the output. This model assumes that the output is a linear function of the parameters and the input.

\noindent $\bullet$ \textbf{Random Forest (\textit{RF})}: A weighted average (ensemble) of multiple Decision Tree classifiers \cite{breiman2001random}. Each decision tree is a non-parametric model based on a series of if-then statements.

\noindent $\bullet$ \textbf{Neural Network (\textit{NN})}: A parametric model trained using the backpropagation algorithm \cite{werbos1974beyond} to minimize a loss function such as cross entropy in the case of classification. The input is passed through a series of weighted linear transforms (similar to Logistic Regression) and non-linear functions such as the hyperbolic tangent or sigmoid function. 

While there have been advances observed using more complex neural network models to encode software artifacts~\cite{xu2016predicting, guo2017semantically}, we exclude them from this work because of the constraints posed by our dataset. Primarily, the highly imbalanced dataset contains many categories that only have a small number of instances (see Table~\ref{table:linkbreakdown}) that are insufficient to train more powerful models without over-fitting.

\begin{table*}[t!]
\centering
\caption{The definition of hyperparameter and the values in the search space. Defaults shown in bold. In the keys, SM = SMOTE, LR = Logistic Regression, RF = random Forest, NN = Neural Network.}
\begin{tabular}{c|l|l}
\toprule
HyperPara. & Definition  &Values                                                     \\
\midrule
SM\_k     & The $k$ nearest neighbours to create synthetic SMOTE examples. & $\{1, 2, 3, 4, \bm{5}, 6, 7\}$                                     \\ 
LR\_c     & Inverse L2 regularization strength: ($\alpha^{-1}$). & $10^{\{-2, -1, \bm{0}, 1, 2\}}$                                  \\ 
RF\_e     & The number of decision tree estimators. & $\{\bm{10}, 100, 1000\}$                                        \\ 
RF\_f     & The function which determines the max. number of features when adding branches. & \{$log_2(|encoding|), \bm{\sqrt{|encoding|}}$\}                 \\
NN\_a     & L2 Regularization strength $\alpha$. & $10^{\{\bm{-4}, -3, -2, -1\}} $                                 \\ 
NN\_dp    & The dropout probability in the hidden layer.  & $\{0.1, 0.25, \bm{0.5}\}$                                     \\ 
NN\_e     & The number of training epochs. & $\{\bm{25}, 50, 75, 100, 125\} $                                \\ 
NN\_lr    & The learning rate used to update the parameters. & $10^{\{\bm{-3}, -2, -1, 0\}} $                              \\ 
\bottomrule
\end{tabular}\\
\label{table:hyperparameter_values}
\end{table*}

\subsection{Model Hyper-parameter Tuning}
\label{subsec:hpram_search}


Table~\ref{table:hyperparameter_values} summarizes the chosen hyper-parameters that are relevant to the classifiers and their values to tune. While this list of hyper-parameters is not exhaustive, preliminary experiments showed this subset to have the greatest effect on overall performance. To balance the cost and potential benefit of tuning, we use a random search strategy to sample a value for each hyper-parameter for a fixed number of iterations~\cite{bergstra2012random}. Additionally, to contend with the large imbalances between link labels discussed in Section~\ref{subsec:all_projects}, we experiment with using the SMOTE~\cite{chawla2002smote} algorithm to create synthetic data samples based on the distribution of nearest neighbors for each existing data point. The parameters of SMOTE are also tuned based on recommendations in related work \cite{agrawal2018better}.

\subsection{Experiment Design}
\label{subsec:semantic_recovery_exp}
We implement the complete link label recovery solution using Python 3.7, Scikit-learn \cite{scikit-learn} and Pytorch \cite{paszke2017automatic}. This solution can be configured with several variations, including choosing the feature-set to encode the links (text features, issue metadata, or both) and the model to perform the actual learning and classification (i.e. LR, RF, or NN). To properly evaluate the performance of our proposed solution on the link label recovery task, we use the cross-validation technique to assess the model performance on the dataset generated in Section~\ref{subsec:case_study_dataset}. This experiment aims to answer \textbf{RQ2}, i.e. to what extent can we recover link labels using machine learning techniques combining the textual content of issues, textual content external to the ITS and the context information about the issues and issue links.

Specifically, we perform 5-fold stratified cross validation to maintain the distribution of the link labels in both the training and testing set. Each time, four folds of data are used for training the model while the remaining one fold is used for testing. During model training, we further split the training data to perform hyper-parameter tuning. On Logistic Regression and Random Forest classifiers, we use 5-fold cross-validation on the training data to find the optimal configurations from the search space in Table \ref{table:hyperparameter_values}. For the neural network, because of the high training cost, we use 25\% of the available training data as a validation set to choose the values for hyper-parameters. 

On the testing data, we use the model output to calculate the F1 measures for each link label. F1 is interpreted as the harmonic mean of the precision and recall, where precision details how many predictions were correct and recall measures how much of the true labels were preserved by the classifier's output. Since the number of instances for different link labels are highly imbalanced, we weigh the F1 measure by the frequency of label occurrences when aggregating the result across all labels. Such process is repeated five times until all the instances in the dataset are tested once. The performance of the model are then averaged among all five folds. 

To benchmark the label recovery performance, we consider two baselines. The first one is a majority classifier which predicts the most common class regardless of the input, also known as a ZeroR classifier. It is useful to serve as a baseline to assess whether more complex classifiers have learned useful features or have over-fit the training data~\cite{hirao2019review}. The second baseline is using the three classifiers without hyper-parameter tuning. Instead, we use the default values corresponding to the values suggested by the implementing library (entries in bold in Table~\ref{table:hyperparameter_values}). In addition, we compare the performance of the classifiers with and without SMOTE.

\subsection{Experiment Results}
\label{subsec:r2_result}
\subsubsection{Effect of issue link representation}
\begin{figure}[!t]
    \centering
    \begin{minipage}{.4\textwidth}
  \centering
  \includegraphics[width=.9\textwidth]{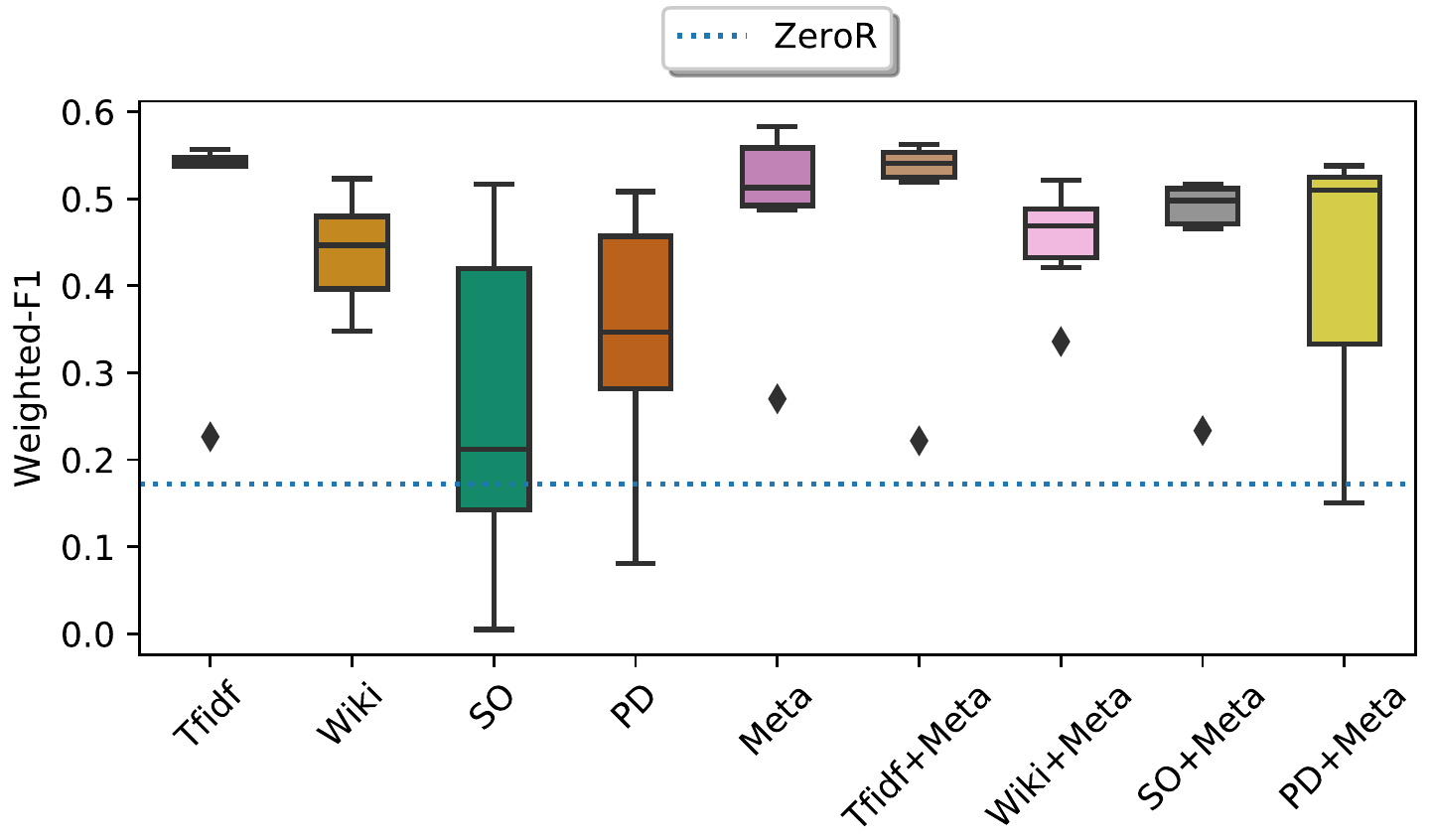}
  \subcaption{F1 obtained with different link representation in Ambari.}
  \label{subfig:ambari_box_plot}
    \end{minipage}%

\vspace{2pt}   
\begin{minipage}{.40\textwidth}
  \centering
  \includegraphics[width=.9\linewidth]{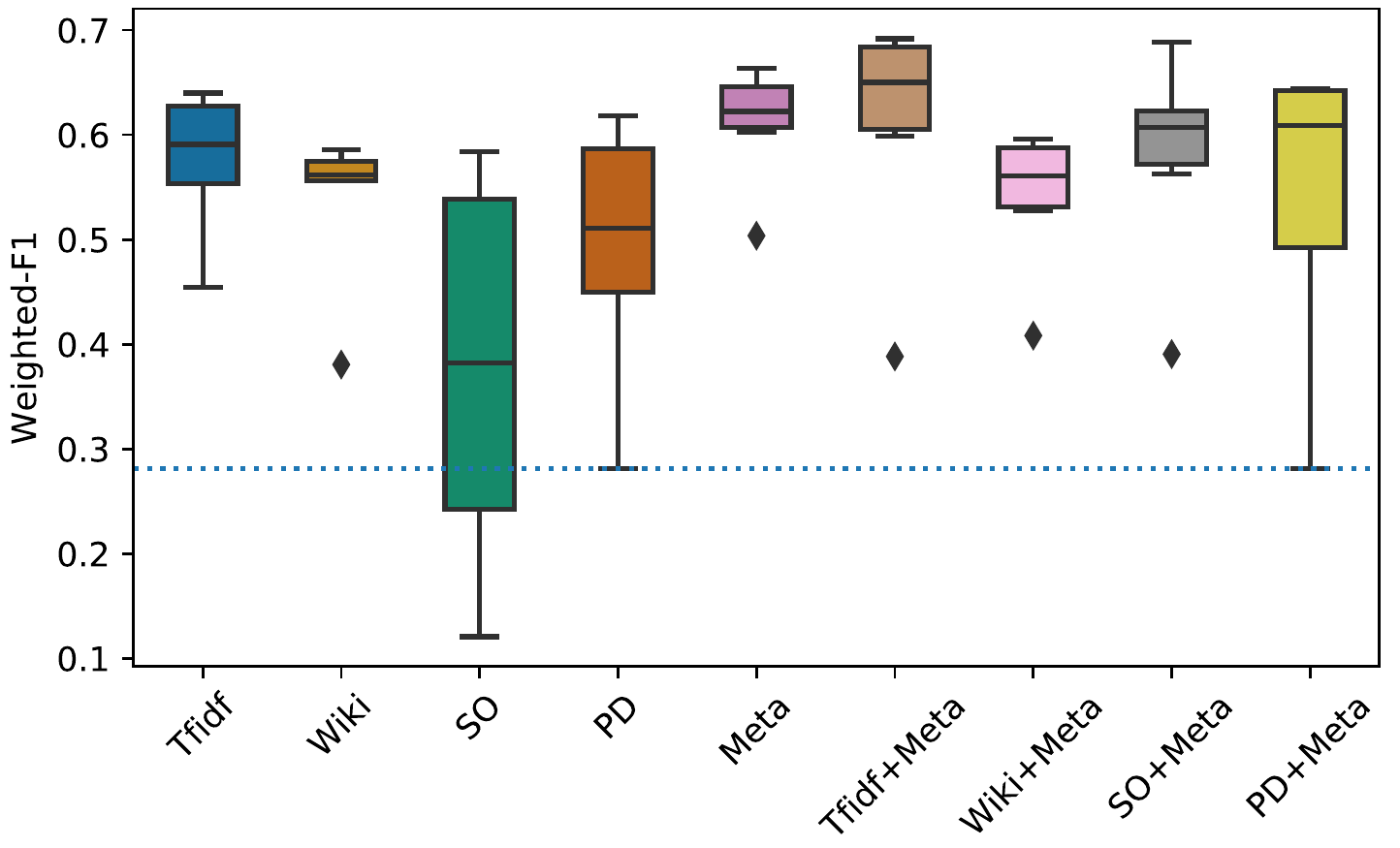}
  \subcaption{F1 obtained with different link representation in Flex.}
  \label{subfig:flex_box_plot}
\end{minipage}

\vspace{2pt}
\begin{minipage}{.40\textwidth}
  \centering
  \includegraphics[width=.9\linewidth]{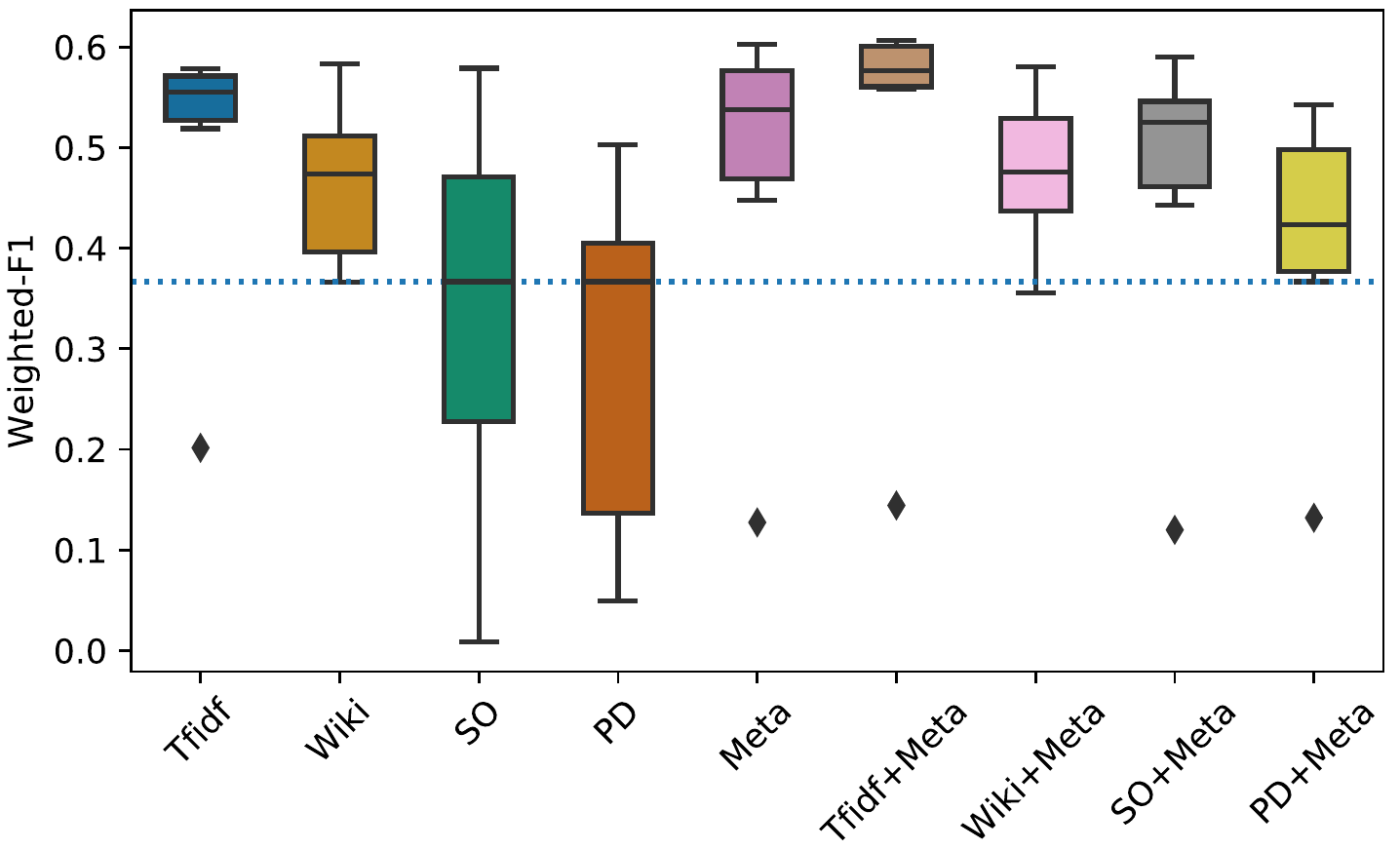}
  \subcaption{F1 obtained with different link representation in Hive.}
  \label{subfig:hive_box_plot}
\end{minipage}

    \caption{Comparison of using different link representation techniques. Wiki=Wikipedia, SO=StackOverflow, PD=Project Documentation.}
    \label{fig:f1_issue_encoding_projects}
\end{figure}

Figure~\ref{fig:f1_issue_encoding_projects} depicts the F1 scores obtained using different link representations. The variance for each link representation comes from the use of SMOTE and the selection of the classifier. The variance is relatively small when using the TF-IDF model to encode the text content of the issues. When using external resources, however, the variance generally increases drastically. Integrating the external resources through word embedding did not improve the performance on the link label recovery task compared to using the simpler, TF-IDF model. The benefit of using metadata, on the other hand, is apparent. All the link representations show increased performance when including additional metadata. The best configuration of link representation is using TFIDF + meta information: the median of weighted F1 measure is 0.54 in Ambari, 0.65 in Flex and 0.58 in Hive.

\subsubsection{Effect of handling imbalance classes}
We decompose the results of using the best link representations from the previous observations. Table~\ref{table:results_tfidf_meta} compares the performance of our method when SMOTE is turned off versus when SMOTE turned on and its parameter is included during model tuning. All the best performance excels ZeroR largely. The effect of having SMOTE, however, varies widely. While SMOTE reduced the performance of the NN model, we observe only a modest change for LR and RF. One potential explanation is that the feature representation used in this experiment is very high-dimensional. Previous work has shown that oversampling high dimensional data is problematic for some machine learning classifiers \cite{Blagus2010}. We leave rigorous investigation on the cause of the limited or even negative effect of using SMOTE to our future work.

\begin{table}[!t]
\begin{center}
\caption{Weighted F1 using the TF-IDF text encoding function with metadata. For each classifier, the performance is break down to without SMOTE (first value in cell) and with SMOTE (second value in cell). The best performance for each dataset is highlighted in red.}
\begin{tabular}{|l|c|c|c|}
\hline
           & Ambari     & Flex     & Hive   \\ \hline
LR         & {\color{red}0.563}, 0.557   & 0.687, 0.675  & {\color{red}0.607}, 0.605 \\ \hline 
NN         & 0.540, 0.222 & {\color{red}0.692}, 0.388 &  0.587, 0.144\\ \hline
RF         & 0.519, 0.540 & 0.599, 0.625  & 0.558, 0.566 \\ \hline
ZeroR      & 0.367  & 0.281 &0.172   \\ \hline 
\end{tabular}
\vspace{-5mm}
\label{table:results_tfidf_meta}
\end{center}
\end{table}

\subsubsection{Effect of projects}
The best models in our method achieved a weighted F1 score of 0.56 in Ambari, 0.70 in Flex, and 0.61 in Hive. Comparing the three projects with various configurations, Hive and Ambari showed much less variance than Flex. This may be explained by the disparity among the size of the dataset. The stability of machine learning techniques depends greatly on the quality and quantity of available training data. Due to the small size of Flex, minute perturbations caused by the randomization process might lead to varied classification results for each label. In such a case, both correct and incorrect classifications on a single label will affect the F1 measure to a greater extent. 

\begin{figure}[!t]
    \centering
    \begin{minipage}{.45\textwidth}
  \centering
  \includegraphics[width=.90\textwidth]{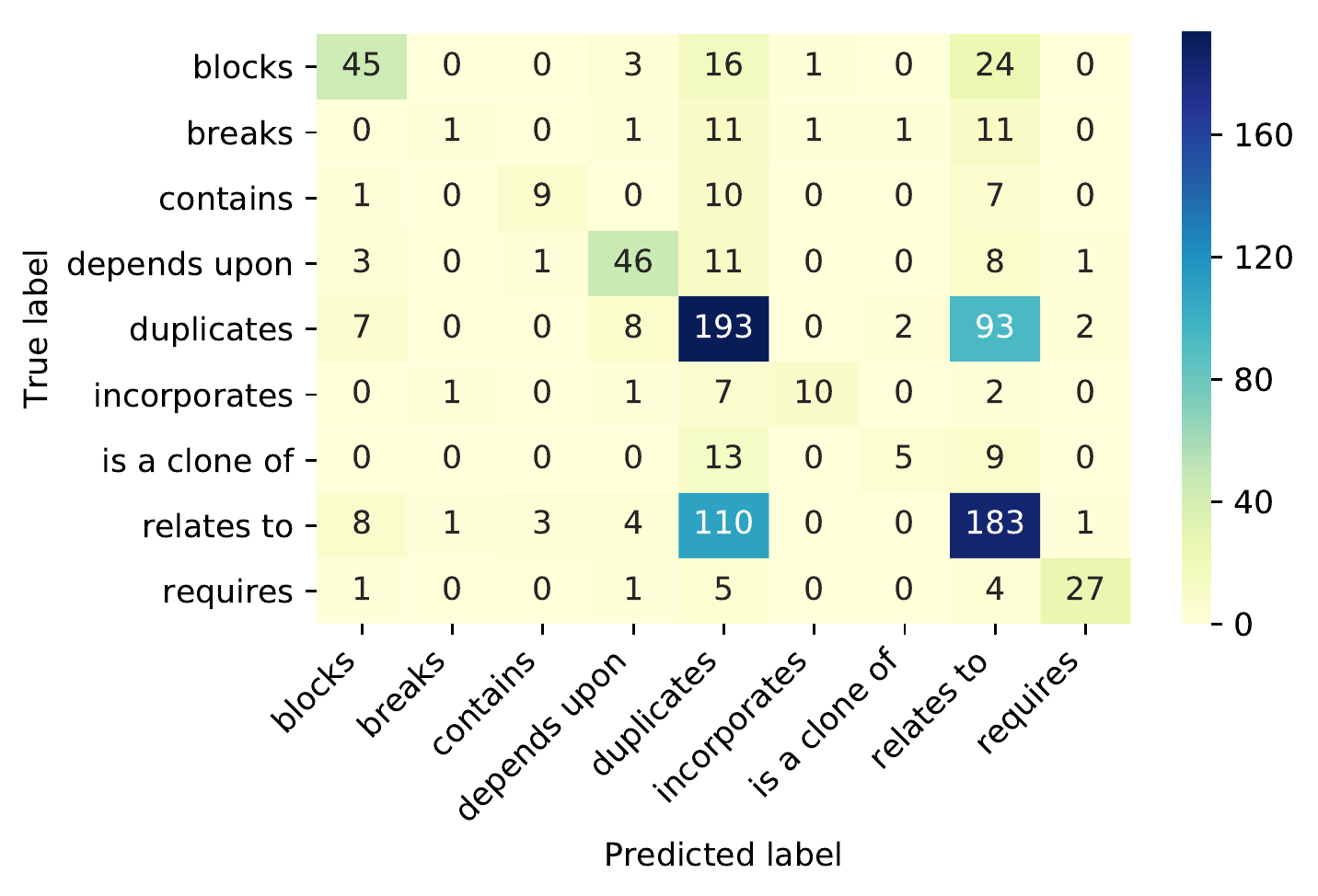}
  \subcaption{Confusion Matrix showing predictions for the best performing configuration (LR, Tfidf, Smote) in Ambari.}
  \label{subfig:ambari_confusion_matrix}
    \end{minipage}%
    
\begin{minipage}{.40\textwidth}
  \centering
  \includegraphics[width=.90\linewidth]{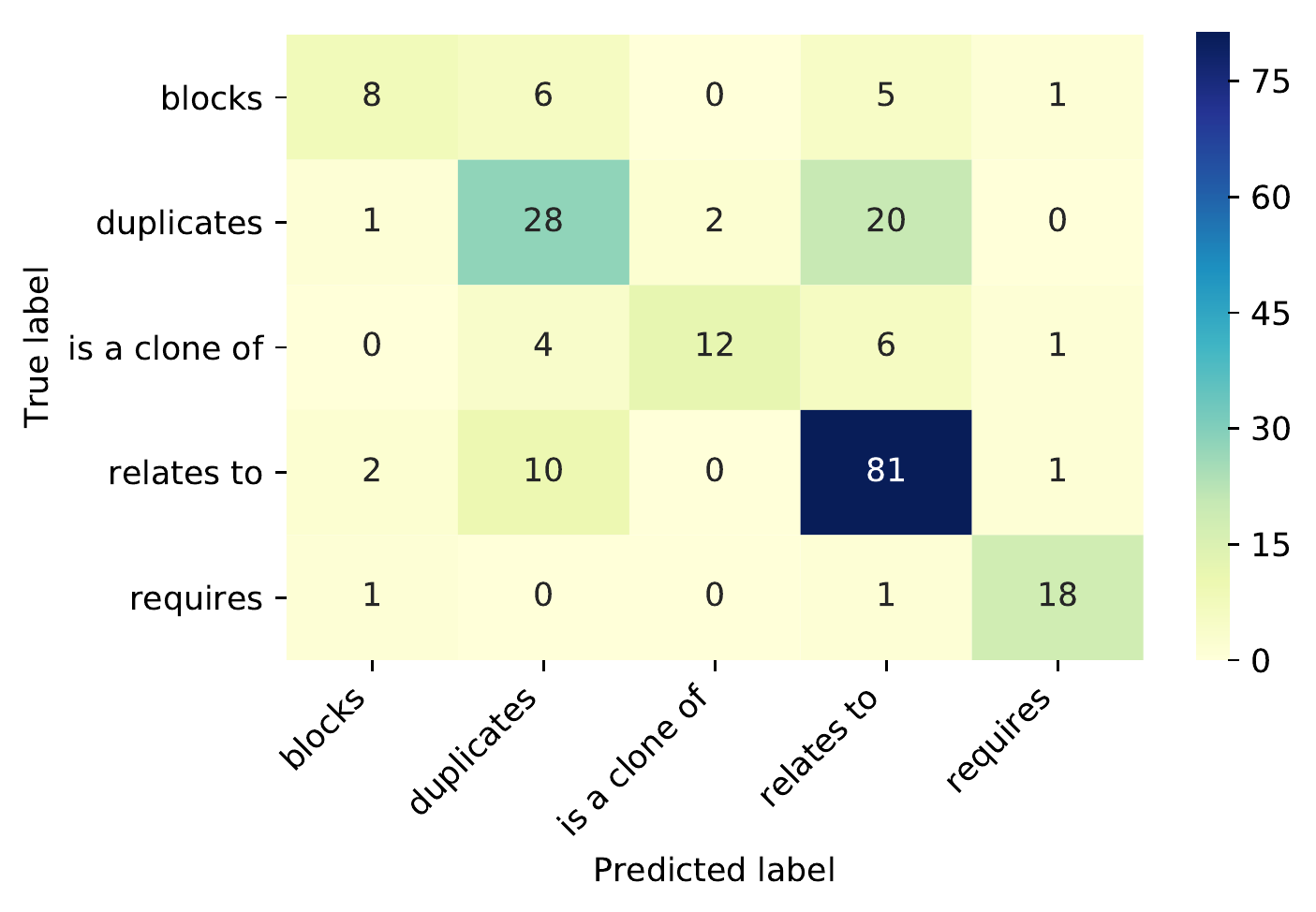}
  \subcaption{Confusion Matrix showing predictions for the best performing configuration (LR, Tfidf+Meta, No Smote) in Flex.}
  \label{subfig:flex_confusion_matrix}
\end{minipage}

\begin{minipage}{.45\textwidth}
  \centering
  \includegraphics[width=.90\linewidth]{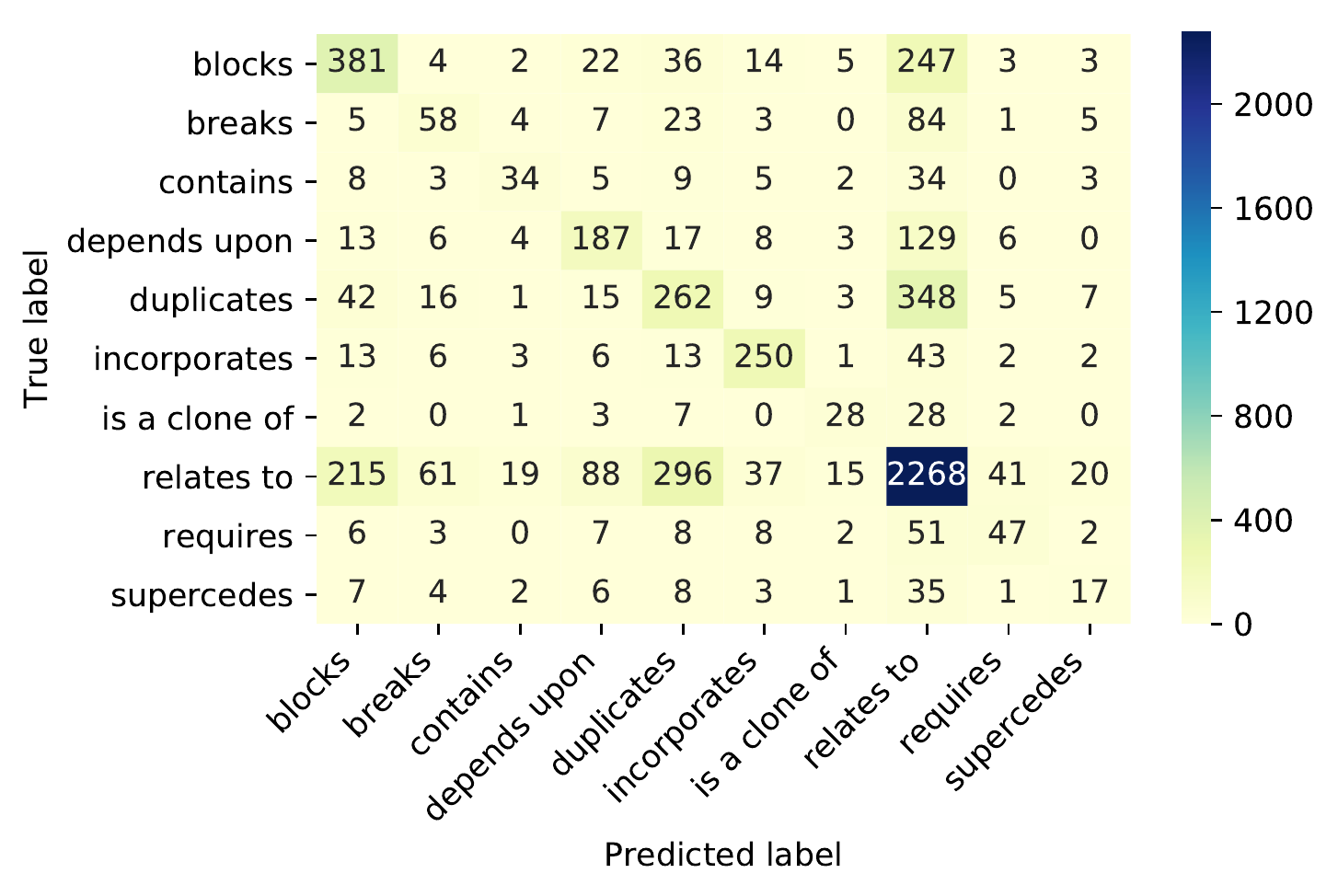}
  \subcaption{Confusion Matrix showing predictions for the best performing configuration (LR, Tfidf+Meta, Smote)
in Hive.}
  \label{subfig:hive_confusion_matrix}
\end{minipage}

    \caption{Confusion Matrices detailing predictions.}
    \label{fig:confusion_matrices_projects}
\end{figure}

Figure \ref{fig:confusion_matrices_projects} shows the confusion matrix of test predictions using the best performing model on each dataset. For project Ambari and Flex, links with various labels can be easily misclassified as \textit{relates to} and \textit{duplicates}. The big number of instances of these two classes in the training set might be the culprit and using SMOTE in Ambari is not particularly effective. In Hive, the classifier is also having difficulty to differentiate \textit{relates to} with other labels. It might also be caused by the lack of a clear definition of the link labels when using the \textit{relates to} label. The links might exhibit heterogeneous features. On the other hand, for some labels, such as \textit{depends upon}, \textit{incorporates} and \textit{requires}, the best models are very effective, achieving an highest F1 score of between 0.69 and 0.88. Such encouraging results indicate the potential of learning accurate link labels to support specific tasks such as feature dependency analysis and work planning.

In summary, the performance of the current link label recovery solution is promising, yet not perfect, especially for the labels with limited instances during training. As a first step, it provides useful information to mitigate the effort of assigning link labels from scratch. The output of the method can suggest link labels in a way that resembles past usage. Such support can also potentially reduce the inconsistency when manually assigning labels.

\section{Predicting Future Link Labels}
\label{sec:semantics_prediction}
In the previous sections, we attempted to model the task of recovering missing links given the entirety of the project context. Another common scenario, however, is predicting future links given a model trained using past links. This scenario cannot be evaluated using stratified cross-validation because the future links and their labels are not available during the model training. We can, however, gain a more accurate picture of prediction performance by varying the training and testing \textit{dates}. In this section, we investigate the extent to which predicting future link labels can be accomplished by our automated solution (our \textbf{RQ3}).

\subsection{Experiment Design}
The complete traceability solution in the prediction setting will involve creating the links first. This can be done either manually by the stakeholders manually or by relying on the vast amount of previous work on automated traceability~\cite{guo2017semantically, heck2014horizontal}. Therefore, the evaluation of this preceding step is out of the scope of this work. We focus instead on evaluating the performance of the next step, predicting the labels when the links are being created. While the models themselves won't be changed from the recovery setting, the performance of the models may vary greatly. Firstly, in the prediction case, the model may encounter word tokens it has never seen during training, known as an Out of Vocabulary (OOV) issue. This can be handled in a number of different ways such as selectively updating the model~\cite{bassiou2014online}, or simply ignoring the new value. In this work for the sake of simplicity, we replace out of vocabulary word tokens by a dummy value that will not change the issue's characterization. Secondly, the distribution of the training data and the testing data might differ depending on the phase in which the project is situated. To capture various adaptations of this scenario, we split the training and testing data in two ways based on the time of issue creation. 

We first find the date at which 60\% of the issues in the project were created. From those issues, we create a training set of links only between these issues. The model tuning and training are performed on this training set. When we prepare the word embeddings, we also use this set to update the embeddings as discussed in Section~\ref{sec:semantics_features} to prevent information about future issues ``leaking'' into the training process. The next 20\% of issues are used to create the test set, i.e. links from these new issues to the complete set of issues. We denote this setting as $60-20$. This process is then repeated using the 80\% date issues as the training set and the last 20\% data issues as the testing set for comparison, denoted by $80-20$. Such a division of training and testing data in two settings aims to make a fair comparison of the performance on test sets.  Similar to Section~\ref{subsec:semantic_recovery_exp}, we perform the hyper-parameter tuning on the classifiers and use the F1 measure for each link label on the held-out future test set.

\subsection{Experiment Result}
As shown in Table~\ref{table:time_general}, the best models achieved a weighted F1 score of 0.326 in Ambari in the setting of $60-20$, 0.570 in Flex, and 0.333 in Hive from the same setting. Compare to the best result achieved in the recovery scenario (0.56 in Ambari, 0.70 in Flex, and 0.61 in Hive), such amount of deterioration can reasonable considering the smaller amount of training data, the out-of-vocabulary issue and label distribution shift as we discussed earlier. However, when the model is trained in the later stage of the project, the performance of the model can improve greatly, ranging between 18\% to 42\%.

\begin{table}[!t]
\begin{center}
\caption{Weighted F1 when predicting future link labels in two time-based evaluation settings, $60-20$ and $80-20$. }
\begin{tabular}{|l|c|c|c|}
\hline
           & Ambari     & Flex     & Hive   \\ \hline
$60-20$        & 0.326 & 0.570 &  0.333 \\ \hline
$80-20$         & 0.416   & 0.669  & 0.416 \\ \hline 
\end{tabular}
\vspace{-5mm}
\label{table:time_general}
\end{center}
\end{table}

\begin{figure}[!t]
    \centering
    \begin{minipage}{.42\textwidth}
  \centering
  \includegraphics[width=.9\textwidth]{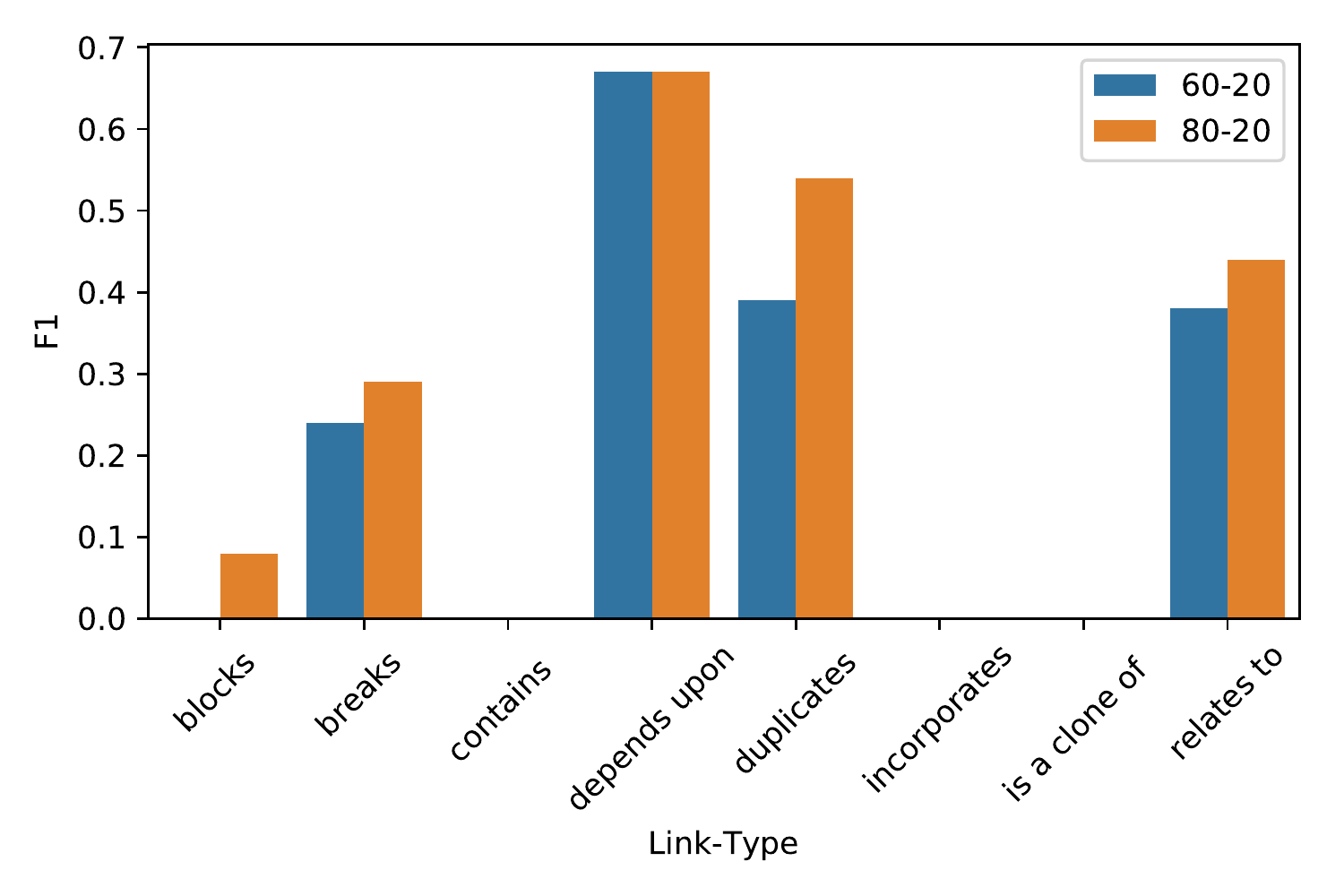}
  \subcaption{F1 measure breakdown in Ambari.}
  \label{subfig:ambari_time}
    \end{minipage}%
    
\begin{minipage}{.42\textwidth}
  \centering
  \includegraphics[width=.9\linewidth]{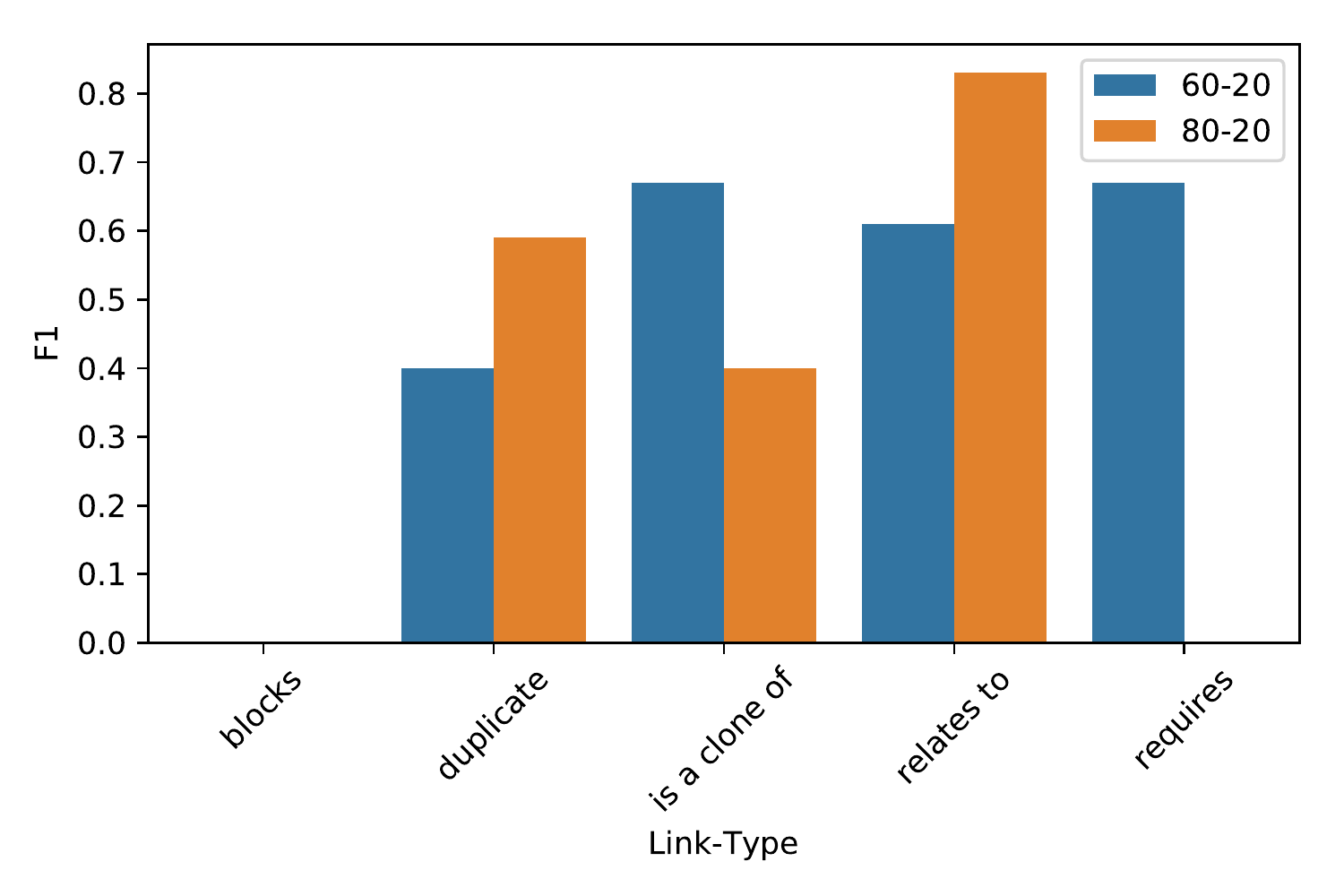}
  \subcaption{F1 measure breakdown in Flex.}
  \label{subfig:flex_time}
\end{minipage}

\begin{minipage}{.42\textwidth}
  \centering
  \includegraphics[width=.9\linewidth]{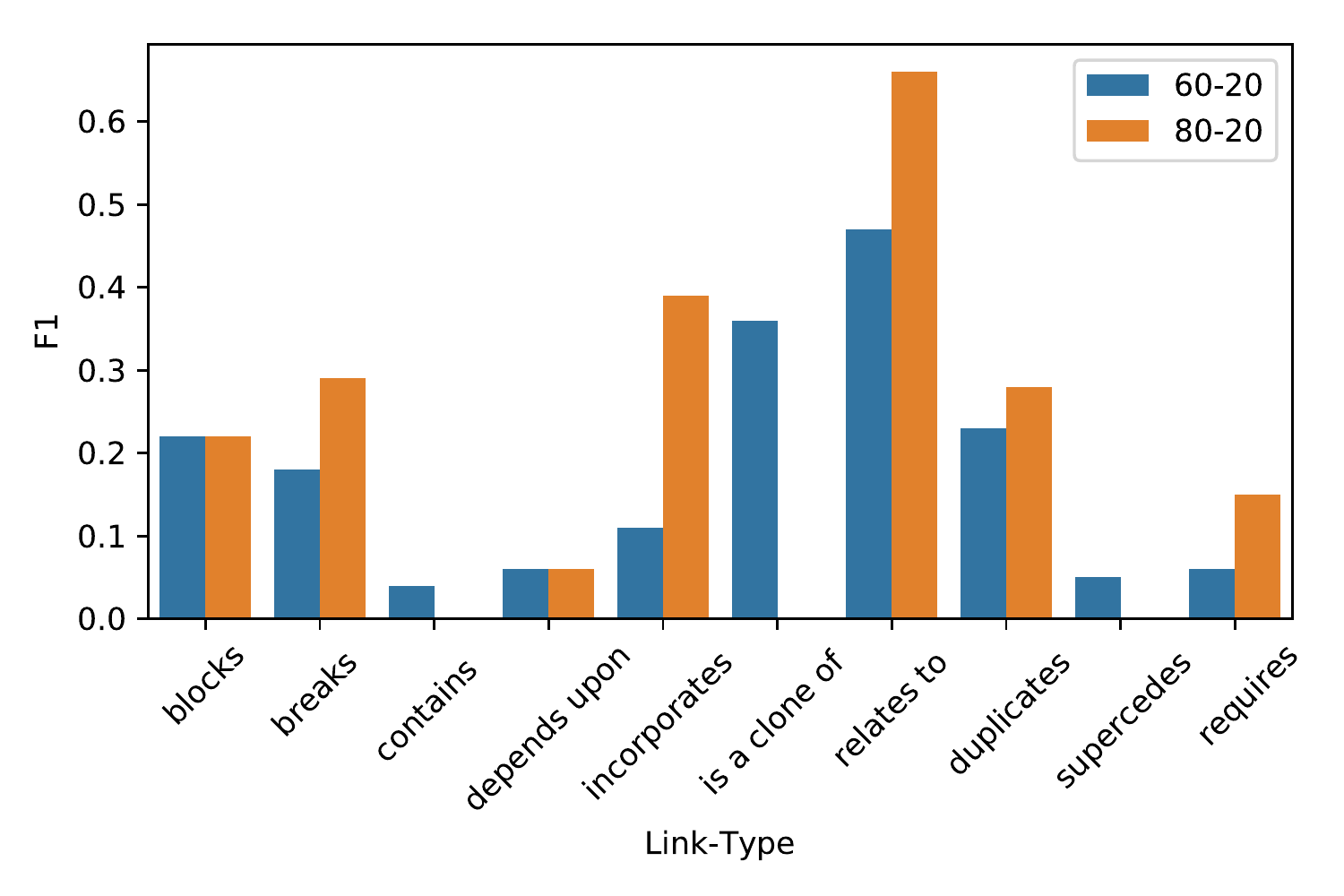}
  \subcaption{F1 measure breakdown in Hive.}
  \label{subfig:hive_time}
\end{minipage}

    \caption{Performance of the link label prediction task}
    \label{fig:time_based_breakdown_projects}
\end{figure}

We breakdown the performance for each label from the test set in Figure~\ref{fig:time_based_breakdown_projects}. The performance of our method varies for different labels, similar to the trend observed in the label recovery task. The missing bar indicates no prediction on that label. Comparing the setting of $60-20$ and $80-20$, the performance of most labels has demonstrated improvement. The exceptions are \textit{requires} in Flex, and several other labels in Flex and Hive. The first case happens because \textit{requires} has not shown in the testing data of $80-20$ so that we cannot estimate our model on such instances. We zoomed in the other cases and found the configuration of the best model changes from using Wiki Embedding to the TfIDF model for Hive when the training data increase from 60\% to 80\%. Because we perform hyper-parameter tuning during training, while the best configuration is performing better in general, it might misclassify a small number of instances to preserve the general trend of improvement. These results indicate that while our solution can perform reasonably well in prediction scenarios, especially when the historical data accumulate, human inspection is still necessary to ensure the high quality of the suggested link labels.

\section{Threats to Validity}
\label{sec:threatsvalidity}

\textit{Internal Validity}: Automated solutions including machine learning often require making a number of related decisions in order to optimize performance. These decisions can include the type of classifier and the subsequent set of hyper-parameters and their values, the feature representation of input values, the evaluation criteria and metrics, and the optimization methods. Evaluating every possible combination of these decisions is at best resource intensive. In this work, we have presented experimental results on a subset of these potential choices based on prevalence in previous work as well as preliminary empirical experimentation. Through careful consideration of issues such as hyper-parameter tuning, handling unbalanced classes, and treating time-sensitive attributes, our study carries direct implications when applying learning-based methods on practical link labeling problems. 


\textit{External Validity}: The datasets used in this work primarily focused on Apache projects and the Jira ITS. Some findings may not transfer to projects that do not use Jira to track issues. Wider studies on link types including additional ITSs would be required to make more general statements about trace link types. We note, however, that even within the Apache ecosystem, trace link usage varies between projects (see Section~\ref{subsec:case_study_dataset}). Intentionally, the label recovery experiments presented in this work do not presuppose specific link labels so the steps presented here should transfer to other datasets.



\section{Conclusions}
\label{sec:conclusion}
In this work, we have investigated horizontal traceability from the perspective of issue link labels in open source projects. Our study indicates that, for projects with varying size and complexity, it is possible to achieve fairly high recovery scores (a weighted F1 measure of 0.56-0.70) for trace link label recovery using machine learning techniques, model tuning, and feature selection. Particularly, we show that the set of features, the classifier, and the hyper-parameter settings presented all greatly influence the performance of the label recovery method and careful consideration of validation criteria and available resources should be made before implementing any single solution. While we haven't observed obvious advantages of pre-training word embeddings in our case study, the positive effect of using metadata of the issue links is decisive. This work also demonstrated the potential of using the machine learning models to predict labels for future links. As the project proceeds, and more training data is available, our method brings improvement on the label prediction performance.

We envision two directions to which this work can directly bring benefits. First, our link label recovery function can be used to support many automated analytic tools. Most of the current automated analytic tools require complete and accurate link labels to aggregate information about certain aspects of the development process. For example, the links related to release planning and stability analysis would be entirely different. The analytic results can be sabotaged if the link labels are incomplete or inaccurate. Our method can be used to support identifying issue links that are relevant to the specific analytic functions. The second direction is to build plugins for ITSs to suggest link labels when the users are creating or inspecting related links using the prediction function. Such suggestions will directly support novice contributors who are not knowledgeable enough about the existing code base or issues to make meaningful and accurate links.




\bibliographystyle{abbrv}
\bibliography{knowledgetraceability}

\begin{thebibliography}{10}

\bibitem{abukwaik2018semi}
H.~Abukwaik, A.~Burger, B.~K. Andam, and T.~Berger.
\newblock Semi-automated feature traceability with embedded annotations.
\newblock In {\em 2018 IEEE International Conference on Software Maintenance
  and Evolution (ICSME)}, pages 529--533. IEEE, 2018.

\bibitem{agrawal2018better}
A.~Agrawal and T.~Menzies.
\newblock Is better data better than better data miners?: on the benefits of
  tuning smote for defect prediction.
\newblock In {\em Proceedings of the 40th International Conference on Software
  Engineering}, pages 1050--1061. ACM, 2018.

\bibitem{arya2019analysis}
D.~Arya, W.~Wang, J.~L. Guo, and J.~Cheng.
\newblock Analysis and detection of information types of open source software
  issue discussions.
\newblock {\em arXiv preprint arXiv:1902.07093}, 2019.

\bibitem{bassiou2014online}
N.~K. Bassiou and C.~L. Kotropoulos.
\newblock Online plsa: Batch updating techniques including out-of-vocabulary
  words.
\newblock {\em IEEE transactions on neural networks and learning systems},
  25(11):1953--1966, 2014.

\bibitem{bergstra2012random}
J.~Bergstra and Y.~Bengio.
\newblock Random search for hyper-parameter optimization.
\newblock {\em Journal of Machine Learning Research}, 13(Feb):281--305, 2012.

\bibitem{Blagus2010}
R.~Blagus and L.~Lusa.
\newblock Class prediction for high-dimensional class-imbalanced data.
\newblock {\em BMC Bioinformatics}, 11(1):523, Oct 2010.

\bibitem{fastText}
P.~Bojanowski, E.~Grave, A.~Joulin, and T.~Mikolov.
\newblock Enriching word vectors with subword information.
\newblock {\em Transactions of the Association for Computational Linguistics},
  5:135--146, 2017.

\bibitem{breiman2001random}
L.~Breiman.
\newblock Random forests.
\newblock {\em Machine learning}, 45(1):5--32, 2001.

\bibitem{chawla2002smote}
N.~V. Chawla, K.~W. Bowyer, L.~O. Hall, and W.~P. Kegelmeyer.
\newblock Smote: synthetic minority over-sampling technique.
\newblock {\em Journal of artificial intelligence research}, 16:321--357, 2002.

\bibitem{P18-1198}
A.~Conneau, G.~Kruszewski, G.~Lample, L.~Barrault, and M.~Baroni.
\newblock What you can cram into a single {\$}{\&}!{\#}* vector: Probing
  sentence embeddings for linguistic properties.
\newblock In {\em Proceedings of the 56th Annual Meeting of the Association for
  Computational Linguistics (Volume 1: Long Papers)}, pages 2126--2136.
  Association for Computational Linguistics, 2018.

\bibitem{da2016impact}
D.~A. da~Costa, S.~McIntosh, U.~Kulesza, and A.~E. Hassan.
\newblock The impact of switching to a rapid release cycle on the integration
  delay of addressed issues-an empirical study of the mozilla firefox project.
\newblock In {\em 2016 IEEE/ACM 13th Working Conference on Mining Software
  Repositories (MSR)}, pages 374--385. IEEE, 2016.

\bibitem{Dahlstedt2005}
{\AA}.~G. Dahlstedt and A.~Persson.
\newblock {\em Requirements Interdependencies: State of the Art and Future
  Challenges}, pages 95--116.
\newblock Springer Berlin Heidelberg, Berlin, Heidelberg, 2005.

\bibitem{Do2017RefinementAR}
A.~Q. Do and T.~Bhowmik.
\newblock Refinement and resolution of just-in-time requirements in open source
  software: A case study.
\newblock {\em 2017 IEEE 25th International Requirements Engineering Conference
  Workshops (REW)}, pages 407--410, 2017.

\bibitem{davide_TSE2018}
D.~Falessi, J.~Roll, J.~L. Guo, and J.~Cleland-Huang.
\newblock Leveraging historical associations between requirements and source
  code to identify impacted classes.
\newblock {\em IEEE Transactions on Software Engineering}, pages 1--1, 2018.

\bibitem{fu2017easy}
W.~Fu and T.~Menzies.
\newblock Easy over hard: A case study on deep learning.
\newblock In {\em Proceedings of the 2017 11th Joint Meeting on Foundations of
  Software Engineering}, pages 49--60. ACM, 2017.

\bibitem{guo2017semantically}
J.~Guo, J.~Cheng, and J.~Cleland-Huang.
\newblock Semantically enhanced software traceability using deep learning
  techniques.
\newblock In {\em Software Engineering (ICSE), 2017 IEEE/ACM 39th International
  Conference on}, pages 3--14. IEEE, 2017.

\bibitem{guo_re2013}
J.~{Guo}, J.~{Cleland-Huang}, and B.~{Berenbach}.
\newblock Foundations for an expert system in domain-specific traceability.
\newblock In {\em 2013 21st IEEE International Requirements Engineering
  Conference (RE)}, pages 42--51, July 2013.

\bibitem{guo2017tackling}
J.~Guo, M.~Gibiec, and J.~Cleland-Huang.
\newblock Tackling the term-mismatch problem in automated trace retrieval.
\newblock {\em Empirical Software Engineering}, 22(3):1103--1142, 2017.

\bibitem{Guo:2014:TID:2642937.2642970}
J.~Guo, N.~Monaikul, C.~Plepel, and J.~Cleland-Huang.
\newblock Towards an intelligent domain-specific traceability solution.
\newblock In {\em Proceedings of the 29th ACM/IEEE International Conference on
  Automated Software Engineering}, ASE '14, pages 755--766, New York, NY, USA,
  2014. ACM.

\bibitem{heck2014horizontal}
P.~Heck and A.~Zaidman.
\newblock Horizontal traceability for just-in-time requirements: the case for
  open source feature requests.
\newblock {\em Journal of Software: Evolution and Process}, 26(12):1280--1296,
  2014.

\bibitem{hirao2019review}
T.~Hirao, S.~McIntosh, A.~Ihara, and K.~Matsumoto.
\newblock The review linkage graph for code review analytics: a recovery
  approach and empirical study.
\newblock In {\em Proceedings of the 2019 27th ACM Joint Meeting on European
  Software Engineering Conference and Symposium on the Foundations of Software
  Engineering}, pages 578--589. ACM, 2019.

\bibitem{hu2018recommending}
D.~Hu, M.~Chen, T.~Wang, J.~Chang, G.~Yin, Y.~Yu, and Y.~Zhang.
\newblock Recommending similar bug reports: A novel approach using document
  embedding model.
\newblock In {\em 2018 25th Asia-Pacific Software Engineering Conference
  (APSEC)}, pages 725--726. IEEE, 2018.

\bibitem{karpathy2015deep}
A.~Karpathy and L.~Fei-Fei.
\newblock Deep visual-semantic alignments for generating image descriptions.
\newblock In {\em Proceedings of the IEEE conference on computer vision and
  pattern recognition}, pages 3128--3137, 2015.

\bibitem{landauer1997solution}
T.~K. Landauer and S.~T. Dumais.
\newblock A solution to plato's problem: The latent semantic analysis theory of
  acquisition, induction, and representation of knowledge.
\newblock {\em Psychological review}, 104(2):211, 1997.

\bibitem{luders2019openreq}
C.~M. L{\"u}ders, M.~Raatikainen, J.~Motger, and W.~Maalej.
\newblock Openreq issue link map: A tool to visualize issue links in jira.
\newblock In {\em 2019 IEEE 27th International Requirements Engineering
  Conference (RE)}, pages 492--493. IEEE, 2019.

\bibitem{maalej2017using}
W.~Maalej, M.~Ellmann, and R.~Robbes.
\newblock Using contexts similarity to predict relationships between tasks.
\newblock {\em Journal of Systems and Software}, 128:267--284, 2017.

\bibitem{Mder2013StrategicTF}
P.~M{\"a}der, P.~L. Jones, Y.~Zhang, and J.~Cleland-Huang.
\newblock Strategic traceability for safety-critical projects.
\newblock {\em IEEE Software}, 30:58--66, 2013.

\bibitem{DBLP:journals/corr/MerityXBS16}
S.~Merity, C.~Xiong, J.~Bradbury, and R.~Socher.
\newblock Pointer sentinel mixture models.
\newblock {\em CoRR}, abs/1609.07843, 2016.

\bibitem{mikolov2013efficient}
T.~Mikolov, K.~Chen, G.~Corrado, and J.~Dean.
\newblock Efficient estimation of word representations in vector space.
\newblock {\em arXiv preprint arXiv:1301.3781}, 2013.

\bibitem{mishra2019use}
S.~Mishra and A.~Sharma.
\newblock On the use of word embeddings for identifying domain specific
  ambiguities in requirements.
\newblock In {\em 2019 IEEE 27th International Requirements Engineering
  Conference Workshops (REW)}, pages 234--240. IEEE, 2019.

\bibitem{nakov2016semeval}
P.~Nakov, A.~Ritter, S.~Rosenthal, F.~Sebastiani, and V.~Stoyanov.
\newblock Semeval-2016 task 4: Sentiment analysis in twitter.
\newblock In {\em Proceedings of the 10th international workshop on semantic
  evaluation (semeval-2016)}, pages 1--18, 2016.

\bibitem{alexander_nicholson_link_dataset}
A.~Nicholson, D.~Arya, and J.~L. Guo.
\newblock Historical issue data of projects on jira, July 2020.

\bibitem{nicholson2020traceability}
A.~Nicholson, D.~M. Arya, and J.~L. Guo.
\newblock Traceability network analysis: A case study of links in issue
  tracking systems.
\newblock In {\em 2020 IEEE Seventh International Workshop on Artificial
  Intelligence for Requirements Engineering (AIRE)}, pages 39--47. IEEE, 2020.

\bibitem{paszke2017automatic}
A.~Paszke, S.~Gross, S.~Chintala, G.~Chanan, E.~Yang, Z.~DeVito, Z.~Lin,
  A.~Desmaison, L.~Antiga, and A.~Lerer.
\newblock Automatic differentiation in pytorch.
\newblock 2017.

\bibitem{scikit-learn}
F.~Pedregosa, G.~Varoquaux, A.~Gramfort, V.~Michel, B.~Thirion, O.~Grisel,
  M.~Blondel, P.~Prettenhofer, R.~Weiss, V.~Dubourg, J.~Vanderplas, A.~Passos,
  D.~Cournapeau, M.~Brucher, M.~Perrot, and E.~Duchesnay.
\newblock Scikit-learn: Machine learning in {P}ython.
\newblock {\em Journal of Machine Learning Research}, 12:2825--2830, 2011.

\bibitem{icse2018_traceability_in_the_wild}
M.~{Rath}, J.~{Rendall}, J.~L.~C. {Guo}, J.~{Cleland-Huang}, and P.~{Mäder}.
\newblock Traceability in the wild: Automatically augmenting incomplete trace
  links.
\newblock In {\em 2018 IEEE/ACM 40th International Conference on Software
  Engineering (ICSE)}, pages 834--845, May 2018.

\bibitem{Rempel2017PreventingDT}
P.~Rempel and P.~M{\"a}der.
\newblock Preventing defects: The impact of requirements traceability
  completeness on software quality.
\newblock {\em IEEE Transactions on Software Engineering}, 43:777--797, 2017.

\bibitem{scandariato2014predicting}
R.~Scandariato, J.~Walden, A.~Hovsepyan, and W.~Joosen.
\newblock Predicting vulnerable software components via text mining.
\newblock {\em IEEE Transactions on Software Engineering}, 40(10):993--1006,
  2014.

\bibitem{spanoudakis2004rule}
G.~Spanoudakis, A.~Zisman, E.~P{\'e}rez-Minana, and P.~Krause.
\newblock Rule-based generation of requirements traceability relations.
\newblock {\em Journal of systems and software}, 72(2):105--127, 2004.

\bibitem{Sthl2016AchievingTI}
D.~St{\aa}hl, K.~Hall{\'e}n, and J.~Bosch.
\newblock Achieving traceability in large scale continuous integration and
  delivery deployment, usage and validation of the eiffel framework.
\newblock {\em Empirical Software Engineering}, 22:967--995, 2016.

\bibitem{Tian:2014:SSW:2591062.2591071}
Y.~Tian, D.~Lo, and J.~Lawall.
\newblock Sewordsim: Software-specific word similarity database.
\newblock In {\em Companion Proceedings of the 36th International Conference on
  Software Engineering}, ICSE Companion 2014, pages 568--571, New York, NY,
  USA, 2014. ACM.

\bibitem{Tomova:2018:UTL:3183440.3195086}
M.~T. Tomova, M.~Rath, and P.~M\"{a}der.
\newblock Use of trace link types in issue tracking systems.
\newblock In {\em Proceedings of the 40th International Conference on Software
  Engineering: Companion Proceeedings}, ICSE '18, pages 181--182, New York, NY,
  USA, 2018. ACM.

\bibitem{werbos1974beyond}
P.~Werbos.
\newblock Beyond regression:" new tools for prediction and analysis in the
  behavioral sciences.
\newblock {\em Ph. D. dissertation, Harvard University}, 1974.

\bibitem{wu2008interpreting_tfidf}
H.~C. Wu, R.~W.~P. Luk, K.~F. Wong, and K.~L. Kwok.
\newblock Interpreting tf-idf term weights as making relevance decisions.
\newblock {\em ACM Transactions on Information Systems (TOIS)}, 26(3):13, 2008.

\bibitem{xia2015elblocker}
X.~Xia, D.~Lo, E.~Shihab, X.~Wang, and X.~Yang.
\newblock Elblocker: Predicting blocking bugs with ensemble imbalance learning.
\newblock {\em Information and Software Technology}, 61:93--106, 2015.

\bibitem{xu2016predicting}
B.~Xu, D.~Ye, Z.~Xing, X.~Xia, G.~Chen, and S.~Li.
\newblock Predicting semantically linkable knowledge in developer online forums
  via convolutional neural network.
\newblock In {\em Proceedings of the 31st IEEE/ACM International Conference on
  Automated Software Engineering}, pages 51--62. ACM, 2016.

\bibitem{yang2016combining}
X.~Yang, D.~Lo, X.~Xia, L.~Bao, and J.~Sun.
\newblock Combining word embedding with information retrieval to recommend
  similar bug reports.
\newblock In {\em 2016 IEEE 27th International Symposium on Software
  Reliability Engineering (ISSRE)}, pages 127--137. IEEE, 2016.

\bibitem{zhang2006multilabel}
M.-L. Zhang and Z.-H. Zhou.
\newblock Multilabel neural networks with applications to functional genomics
  and text categorization.
\newblock {\em IEEE transactions on Knowledge and Data Engineering},
  18(10):1338--1351, 2006.

\end{thebibliography}

\end{document}